\documentstyle[twocolumn,epsf,rotate,floats,prl,aps]{revtex}
\draft

\begin{document}

\wideabs{
\title{Highly site-specific H$_2$ adsorption on vicinal Si(001) surfaces}
\author{P. Kratzer$^{1}$, E. Pehlke$^{2}$, M. Scheffler$^{1}$}
\address{$^{1}$Fritz-Haber-Institut der Max-Planck-Gesellschaft, 
Faradayweg 4--6, D-14195 Berlin, Germany} 
\address{$^{2}$Physik Department, Technische Universit\"at M\"unchen, 
D-85747 Garching, Germany }

\author{M. B. Raschke and U. H{\"o}fer}
\address{Max-Planck-Institut f{\"u}r Quantenoptik, D-85740 Garching,
Germany 
and Physik Department, Technische Universit\"at M\"unchen, D-85747
Garching, Germany }

\maketitle

\begin{abstract}
Experimental and theoretical results for the dissociative adsorption of 
H$_2$ on vicinal Si(001) surfaces are presented.
Using optical second-harmonic generation, sticking probabilities at 
the step sites are found to exceed those on the terraces by up to six 
orders of magnitude.
Density functional theory calculations indicate the presence of direct
adsorption pathways for monohydride formation but with a 
dramatically lowered barrier for step adsorption due to an efficient 
rehybridization of dangling orbitals.
\end{abstract}


}


In most technological applications of surface chemistry, e.g. in
catalysis, the surfaces used to promote a reaction are highly non-ideal.
They contain steps and other imperfections in large concentrations,
which are thought to provide reactive sites. 
Also in thin-film growth, steps are crucial for producing smooth layers
via the so-called step-flow mode of growth. 
Despite the importance, detailed information about the role of steps is
scarce. 
On metals, it is generally argued that the reactivity at steps is increased
due to a lower coordination number of atoms \cite{Somorjai}.
On semiconductors, the situation is less clear since step and terrace
atoms often attain similar coordination due to special reconstructions. 
The H$_2$/Si system provides a good model to study the role of steps,
since it is the most thoroughly studied adsorption system on a
semiconductor surface and it is of considerable technological relevance
\cite{Chabal}. 
In addition, several recent studies came to conclude that
the interaction of molecular hydrogen would be largely determined by
defect sites \cite{NaJo94,RaCa96b}, and in particular by steps
\cite{JiLu93,HaHa96}. 

In this Letter, we demonstrate that the contributions from terraces and
steps to H$_2$ adsorption on vicinal Si(001) surfaces can be
discriminated using the second harmonic generation (SHG) probe 
technique to monitor hydrogen
coverages during gas exposure.  
The measured sticking coefficients differ by up to six orders of
magnitude and indicate the presence of an efficient adsorption 
pathway at the steps, while adsorption on the terraces involves
a large barrier. 
We performed density functional theory calculations to identify the relevant
reaction mechanisms and to compare their energetics.
Surprisingly, H$_2$ dissociation at the threefold-coordinated 
step atoms is found to proceed via two neighboring sites and 
directly leads to monohydride formation similar as at the dimerized
terrace atoms.
The huge differences in barrier heights arise from the interplay of
lattice deformation and electronic structure effects.
Thus, adsorption on a semiconductor surface may be highly site-specific,
even if the reactive surface atoms have similar coordination.


\begin{figure}[b!]
\leavevmode
\begin{center}
\epsfysize=2.5cm
\epsffile{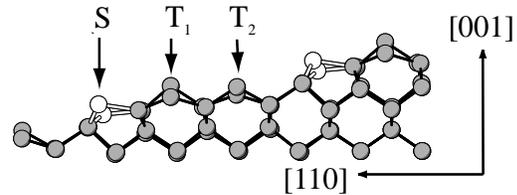}
\end{center}
\caption[]{
Relaxed geometry for the rebonded D$_{\rm B}$ step of a Si(117) surface. 
The rebonded Si atoms are shown in white.
}
\label{geo1}
\end{figure}

The experiments were performed with Si samples that were
inclined from the [001] surface normal towards the [110] direction by
$2.5^\circ$, $5.5^\circ$ and $10^\circ \pm0.5^\circ$. 
They were prepared by removing the oxide layer of the 10-$\Omega$cm
$n$-type wafers in ultra-high vacuum 
at 1250 K followed by slow cooling to below 600 K.
Under these conditions double-atomic height 
steps prevail on the surface
that have additional three-fold coordinated Si atoms attached
to the step edges, the so-called rebonded D$_{\rm B}$ steps\cite{dbsteps}
(a stick-and-ball model is displayed in Fig.~\ref{geo1}).
Low energy electron diffraction (LEED) confirmed that the
surfaces predominantly consisted of a regular array of double-height
steps separated by terraces with a unique orientation of Si dimers. 
The sticking coefficients for the dissociative adsorption of H$_2$ on
these surfaces were determined from the temporal evolution of the 
hydrogen coverage determined during gas exposure by SHG as
described previously \cite{BrKo96}.
Accurate measurements of the H$_2$ desorption process \cite{Hofer92}
ensured that the recorded signal changes were not affected by small
amounts of contaminants in the dosing gas which was purified in liquid
nitrogen traps \cite{BrKo96}.
For the sensitive detection of step adsorption it was exploited that
the presence of regular steps is associated with a symmetry break
in the surface plane. 
For electric field components perpendicular to the step edges this enhances
the SHG contribution of the step with respect to the terrace sites \cite{RaHo98}.

\begin{figure}[t]
\leavevmode
\begin{center}
\epsfxsize=8.0cm
\epsffile{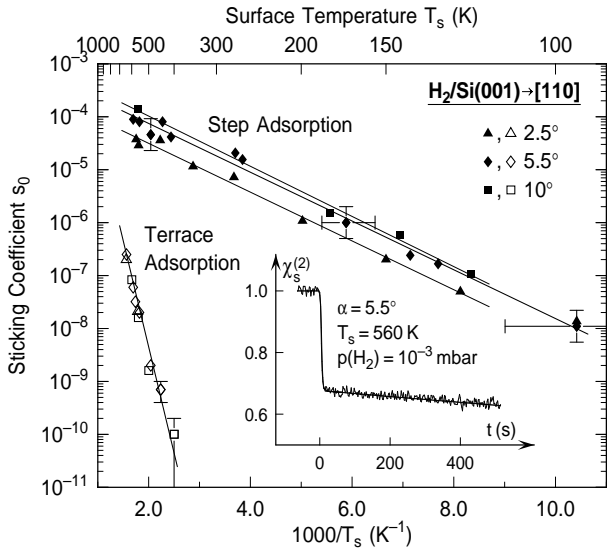}
\end{center}
\caption[]{\label{Arrh} 
Initial sticking coefficients $s_0$ for a gas of H$_2$ at room 
temperature on the steps (filled symbols) and terraces (open symbols) of
vicinal Si(001) surfaces at various surface temperatures $T_s$. 
They were derived from the decay of the nonlinear susceptibility  $\chi^{(2)}_s$
during H$_2$ exposure as shown in the inset. 
Numerical fits to an Arrhenius law $s_0(T_s) = A\exp(-E_a/kT_s)$
yield activation energies $E_a$ and prefactors $A$ for step (terrace)
adsorption of $0.09\pm0.01\,\rm eV$ ($0.76\pm0.05\,\rm eV$) and 
$4\pm2\times10^{-4}$ ($\sim 10^{-1}$).
} 
\end{figure}

A representative measurement taken at the 5.5$^\circ$ sample, kept at
a temperature of 560 K and exposed to H$_2$ at a pressure of 
10$^{-3}$ mbar is displayed in the inset of Fig.~\ref{Arrh}.
There is a rapid drop of the surface nonlinear susceptibility  $\chi^{(2)}_s$
responsible for the SHG signal immediately after admitting H$_2$ gas to
the chamber followed by a more gradual decay. 
The two slopes of the  $\chi^{(2)}_s$ correspond to sticking probabilities of
$1\times 10^{-4}$ and $1.4\times10^{-8}$.
The initial sticking coefficients $s_0$ measured for the different 
samples at various temperatures are collected in the form of two
Arrhenius plots in the main part of Fig.~\ref{Arrh}. 
They span a very wide range from $10^{-10}$ up to $10^{-4}$ \cite{compare}.

We attribute the fast hydrogen uptake -- which is not present on 
the Si(001) surfaces -- to adsorption at special dissociation 
sites of the stepped surface and identify the slow signal decay with
adsorption on terrace sites. 
This interpretation is corroborated by the good agreement between the 
absolute values of the smaller sticking coefficients with those
obtained previously for nominally flat Si(001) \cite{BrKo96} and
by the correlation of the saturation coverage 
$\theta_{\rm step}^{\rm sat}$ associated with the fast decay with the
miscut angle (Table~\ref{satcover}).
$\theta_{\rm step}^{\rm sat}$ was determined by means of temperature
programmed desorption (TPD) after saturation was detected by SHG.  
With the exception of the $10^\circ$ sample 
-- where the number of D$_{\rm B}$ steps might be reduced as a 
result of facetting \cite{Ranke} --
$\theta_{\rm step}^{\rm sat}$ is found to be in good agreement with the
fraction of Si dangling bonds located at the steps relative to the total
amount of dangling bonds in a unit cell of the vicinal surface $R_{\rm db}$
(Table~\ref{satcover}).
Thus it is tempting to associate the hydrogen species adsorbed at the
steps with monohydrides formed by attaching hydrogen to the rebonded Si
atoms. 

For a quantitative comparison of the measured sticking coefficients 
for step and terrace adsorption it is important to know whether they
are due to independent processes.
Annealing and readsorption experiments show that surface temperatures in
excess of 600 K are required to cause appreciable depletion of the
step sites by hydrogen migration on a timescale of several hundred 
seconds \cite{RaHo98}. 
For this reason it can be excluded that hydrogen adsorption on terraces
is mediated by the step sites under the conditions of our experiments. 
The two sticking coefficients given in Fig.~\ref{Arrh} are thus a
quantitative measure of the reactivity of different surface sites. 
The strongly activated behavior observed for terrace adsorption,  
characterized by an Arrhenius energy of $E_a = 0.76\,\rm eV$, is similar
to that reported previously for the well oriented Si(001) and Si(111)
surfaces.
It indicates that distortions of the lattice structure have a pronounced
effect in promoting dissociative adsorption of H$_2$ \cite{BrBr96}.
With $E_a = 0.09\,\rm eV$ the effect of temperature on step 
adsorption is comparatively moderate.

\begin{table}
\begin{tabular}{ccc}
 $\alpha $ & $R_{\rm db}$ & $\theta_{\rm step}^{\rm sat}$ \\ 
\hline 
 2.5$^{\circ}$ & 0.064 & 0.07 \\
 5.5$^{\circ}$ & 0.146 & 0.12 \\
 10$^{\circ}$  & 0.285 & 0.15  \\
\end{tabular}
\medskip
\caption[]{\label{satcover} 
Ratio $R_{\rm db}$ of dangling bonds at rebonded D$_{\rm B}$ steps to total 
number of dangling bonds on vicinal Si(001) surfaces with miscut angle 
$\alpha$ towards [110] and measured saturation coverage of hydrogen at
the steps $\theta_{\rm step}^{\rm sat}$. 
}
\end{table}


The experimental data suggest that the peculiar geometric and electronic
structure of the stepped surface gives rise to reaction channels
that are much more effective than those on ideal surfaces.
To gain an atomistic understanding of the underlying mechanisms we 
determined the total energy of various atomic configurations from
electronic structure calculations. 
We use density functional theory  
with a plane-wave basis set \cite{BoKl97}. The exchange-correlation functional
is treated by the generalized-gradient approximation (GGA) \cite{PeCh92}. 
For silicon we generate an {\em ab initio}, norm-conserving pseudopotential
\cite{Hama89},   
while the bare Coulomb potential is used for hydrogen.
We perform a transition
state search for H$_2$ dissociation in the configuration space of the H atoms and the Si atoms of the
four topmost layers using 
the ridge method \cite{IoCa93}. All calculations, including geometry
optimizations, are performed with 
plane waves up to a cut-off energy of 30 Ry in the basis set. For the
barrier heights reported below, the calculations are repeated with the same geometries, but with a cut-off of 50 Ry.
We model the D$_{\rm B}$ step by a vicinal surface with Miller indices $(1 \, 1 \, 7)$, using a 
monoclinic supercell. 
In this geometry,  
periodically repeated rebonded D$_{\rm B}$ steps are separated by terraces 
two Si dimers wide. Two special {\bf k}-points in the irreducible part of the 
Brillouin zone are used for {\bf k}-space integration. 
The uncertainty in chemisorption energies
due to the finite cell size is determined to be less than 30 meV.

The optimized geometry for the rebonded D$_{\rm B}$ step is shown in Fig.~\ref{geo1}.
The rebonded Si atoms form unusually long bonds to
the atoms at the step edge (mean bond strain 6\%). 
Furthermore, the height of the rebonded Si atoms at the step edge is 
different by 0.67{\AA}, due to a
Jahn-Teller like distortion similar in physical origin to the 
buckling of the Si dimers on the Si(001) surface. 
Consequently, the surface is semiconducting with a Kohn-Sham gap 
of $\sim0.5$eV.  

We investigated the following mechanisms of 
H$_2$ dissociation close to a D$_{\rm B}$ step:
i) Si dimers at the end of a terrace could have different reactivity
compared to the Si dimers on flat regions of the surface.
ii) The H$_2$ breaks the stretched bond of a rebonded Si atom forming a
dihydride species. 
iii) The H$_2$ molecule approaches with orientation parallel to the step
and dissociates, each of the H atoms attaching to one of the Si
rebonded atoms. 

To determine the importance of mechanism i), we locate the
transition states both for H$_2$ dissociation on the terrace Si dimer 
directly above and below the step (T$_1$ and T$_2$ in Fig.~\ref{geo1}).
The geometries of these transition states are asymmetric, with the H$_2$
molecule dissociating above the lower atom of the Si dimer, similar to the 
transition state found for the ideal Si(001) surface \cite{KrHa95}. 
For the barrier heights we obtain 0.40 eV and 0.54 eV for the sites T$_1$ and T$_2$, respectively.
These results are close to the adsorption barrier of 0.4 eV determined 
previously for the flat Si(001) surface using the same
exchange-correlation functional \cite{GrBo97}.
Hence, the Si dimers near steps are only slightly different in reactivity  
from Si dimers on an ideal Si(001) surface.
Thus mechanism i) cannot explain the enhanced reactivity.

Mechanism ii), the formation of a dihydride at the step from a gas phase H$_2$
molecule and a rebonded Si atom, is considerably less exothermic 
(0.9 eV) than monohydride formation ($\sim$2 eV), because the Si--Si 
bond between the rebonded Si atom and the step must be broken.
Nevertheless, dihydrides would exist as metastable species at the 
surface temperatures of the experiment provided the corresponding 
adsorpion barrier were sufficiently low.
However, the calculations yield a barrier of 0.50 eV, even slightly
higher than the one for monohydride formation on the terraces, and
we rule out mechanism ii) as well. 

\begin{figure}
\leavevmode
\begin{center}
\epsfxsize=7.2cm
\epsffile{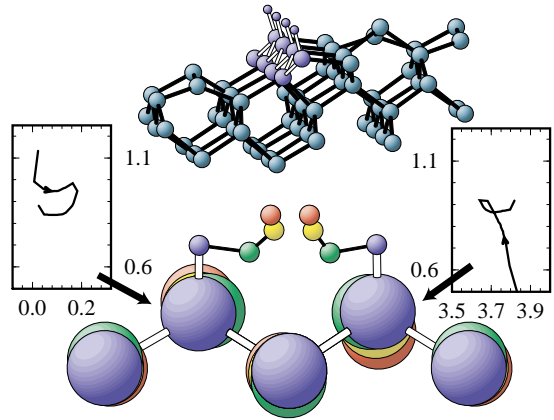}
\end{center}
\medskip
\caption[]{\label{balls} 
Monohydride at the step (upper part) and and motion of the highlighted step atoms during H$_2$ dissociation (lower part).
The reaction path is projected
onto a (110) plane parallel to the D$_{\rm B}$ step with 
different stages of
dissociation marked by different colors.  
The insets show the motion of the rebonded Si atoms (coordinates in {\AA}).  
}
\end{figure}

\begin{table}[b]
\begin{tabular}{lrr}
 site  & $E_{\rm ads}$[eV] & $E_{\rm chem}$[eV] \\ 
\hline
 S (monohydride) & no barrier & $-$2.07 \\
 S (dihydride) & 0.50 & $-$0.87 \\
 T$_1$ & 0.40 & $-$1.75 \\
 T$_2$ & 0.54 & $-$1.93 \\
\end{tabular}
\medskip
\caption[]{\label{Tab2} Adsorption barriers $E_{\rm ads}$ and chemisorption
energies  $E_{\rm chem}$ for H$_2$ 
molecules reacting with the vicinal surface at the sites S, T$_1$ and T$_2$,
as indicated in Fig. \protect\ref{geo1}. 
}
\end{table}
For mechanism iii), monohydride formation from a molecule 
approaching parallel to the step, 
we do not find any barrier. 
Using damped {\em ab initio} molecular dynamics for a slowly approaching
molecule, 
we can identify the reaction path
connecting the gas phase continuously with the monohydride at the step.
Hence we attribute the highly increased sticking coefficient of H$_2$ 
observed on the vicinal surfaces to direct monohydride formation.
This conclusion is confirmed by the observation that this mechanism is
compatible with the observed saturation coverages of
Tab.~\ref{satcover}, as opposed to a complete decoration of the step
with dihydride species, which would result in a saturation coverage a
factor of two higher than observed.
Tab.~\ref{Tab2} summarizes the energetics of the three reaction
mechanisms considered. 

\begin{figure}
\leavevmode
\begin{center}
\epsfysize=5.0cm
\epsffile{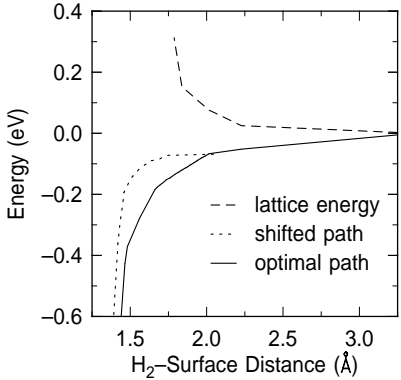}
\end{center}
\caption[]{\label{path} 
 Total energy of the H$_2$/surface system 
 along the adsorption path shown in Fig.~\protect{\ref{balls}} (full line).  
 The dashed line comes from a separate calculation for the bare Si surface 
 using the Si coordinates along the reaction path.
 The thin dotted line denotes the total energy along a similar reaction 
 path, with the two H-adsorption sites translated by one 
 surface lattice constant along the step edge.
 }
\end{figure}

Fig.~\ref{balls} illustrates the 
concerted motion of the H atoms and the two rebonded Si atoms along the
reaction path. 
The Jahn-Teller like splitting 
showing up in the different height of the two rebonded Si atoms is undone 
during the approach of the H$_2$ molecule.
Additionally, the two rebonded Si atoms move closer together by about 
\hbox{0.4{\AA}} during adsorption to assist in breaking the H--H bond.
Both for this optimum pathway and for the similar, 
but slightly less favorable, pathway with the adsorption site 
shifted by one
surface lattice constant parallel to the step, the total energy 
decreases monotonically 
when the H$_2$ molecule approaches the surface (Fig.~\ref{path}, full and
dotted lines). 
This is to be compared to the adsorption energy barrier of 0.3 eV
we have calculated for a rigid Si substrate.
Obviously, a particular distortion of the step Si atoms is crucial 
for adsorption, which, on the bare surface, would be associated 
with a sizeable elastic energy (dashed curve in Fig.~\ref{path}).
Together with the small density of step sites (cf. $R_{\rm db}$ in
Tab.~\ref{satcover}) and 
the necessary alignment of the H$_2$ axis 
parallel to the step,
this explains the small prefactor $A_{\rm step}$
in comparison to $A_{\rm terrace}$. 
The reaction path shown in Fig.~\ref{balls} implies some transfer of energy and
momentum from the H$_2$ to the lattice 
during the adsorption process, which is not easily achieved due to the large 
mass mismatch.
However, thermal fluctuations will sometimes lead to lattice configurations
favorable for adsorption. 
Therefore, the surface temperature dependence of sticking,
as  suggested by the experiments, appears to be compatible with  
the presence of a non-activated adiabatic pathway.  

We find that atomic relaxation 
close to a step, while inducing only moderate changes in the
chemisorption energy \cite{PeKr98}, has a
pronounced influence on the energetics of adsorption. 
Since the early stages of H$_2$ dissociation are quite sensitive to 
electronic states around the Fermi energy, we propose that 
the increase in reactivity is due to shifts of electronic states in
the gap induced by lattice distortions. 
In the surface ground state, the two surface bands
formed from the dangling orbitals located at the 
rebonded Si atoms bracket the Fermi energy and are split by $\sim1$ eV due to 
the Jahn-Teller mechanism.  
However, when the two rebonded
Si atoms are forced to the same geometric height, the energy separation of the 
centers of the surface bands is reduced to 0.4 eV. 
Upon the approach of molecular hydrogen, these
surface states can rehybridize and thus interact efficiently with the H$_2$
molecular orbitals. 
The electronic structure at the step is different from the surface band 
structure of the ideal Si(001) dimer reconstruction: 
At the Si dimers characteristic for the ideal 
surface, the $\pi$-interaction of the
dangling bonds prevents the two band centers to come closer than 0.7eV
\cite{DaSc92}, 
while the dangling bonds of equivalent step edge Si atoms are almost
degenerate. 
Therefore the terrace sites are less capable of dissociating a H$_2$ molecule than the step sites.  

Valuable commentaries and support by  
W. Brenig, K.~L. Kompa and P. Vogl are gratefully acknowledged.
This work was supported in part by the SFB 338 of DFG. 


\begin{thebibliography}{10}
\bibitem{compare}
The obtained values for $s_{0,\rm{step}}$ are not directly comparable with the
results of a previous SHG study of H$_2$ adsorption on vicinal Si surfaces
\cite{HaHa96} 
that reported sticking coefficients as high as 0.1 for
different misorientations and higher surface temperatures.
\end{thebibliography}
\end{document}